\documentclass[jcp, 11pt, notitlepage,amsmath,amssymb]{revtex4-1}

\usepackage{bm,hyperref,tikz,graphicx,calligra}
\usepackage[utf8]{inputenc} 

\begin{document}

\title{A simple model for exploring the role of quantum coherence and the environment in excitonic energy transfer}

\author{Sreenath K.~Manikandan}
\email{sreenath@iisertvm.ac.in}
\affiliation{School of Physics, Indian Institute of Science Education and Research Thiruvananthapuram, CET Campus, Sreekaryam, Thiruvananthapuram, Kerala, India 695016}

\author{Anil Shaji}
\email{shaji@iisertvm.ac.in}
\affiliation{School of Physics, Indian Institute of Science Education and Research Thiruvananthapuram, CET Campus, Sreekaryam, Thiruvananthapuram, Kerala, India 695016}

\begin{abstract}
We investigate the role of quantum coherence in modulating the energy transfer rate between two independent energy donors and a single acceptor participating in an excitonic energy transfer process. The  energy transfer rate  depends explicitly on the nature of the initial coherent superposition state of the two donors and we connect it to the observed absorption profile of the acceptor and the stimulated emission profile of the energy donors. We consider simple models with mesoscopic environments interacting with the donors and the acceptor and compare the expression we obtained for the energy transfer rate with the results of  numerical integration. 
\end{abstract}



\maketitle

\section{Introduction}
The success and applicability of F\"{o}ster's theory of resonant energy transfer~\cite{fster3,fster1} lies in  connecting the expression for the rate of energy transfer between a donor molecule and acceptor to readily measurable spectra of either chromophore. F\"{o}ster Resonant Energy Transfer (FRET)~\cite{fretbook} has been used to understand a wide variety of phenomena starting from the quenching of fluorescence in concentrated dyes~\cite{old1,old2,old3,old4,review} to modeling the efficient energy transfer processes in biological systems~\cite{molenrgy,review,fretbook} including the important problem of understanding the energy harvesting and transfer mechanism in photosynthesis~\cite{psbook,lambert_quantum_2013,fassioli_photosynthetic_2014}. However the theory of FRET was constructed with the energy transfer between a single energy donor and a single acceptor in mind and so one needs to be mindful of this limitation in extending the applicability of the theory to much more complex scenarios like the ones typically encountered in biological systems. Significant progress has been made in recent years both in generalising FRET to more complex system as well as in formulating alternate ways of addressing the problem of energy transfer in biologically relevant systems as can be seen from \cite{Ye:JChemPhys:2012,Renger:PhysChemChemPhys:2013,OlayaCastroScholes2011,Mohseni:JChemPhys:2014,Jang:PhysRevLett:2014,Jang:JPhysChemB:2007,Jang:JChemPhys:2011,Jang:JChemPhys:2009, MC, Jang:WiresComputMolSci:2012}  and references therein. Notably, in \cite{MC} FRET was generalised to the case where there are multiple donors and acceptors. In this Paper we build on the results in \cite{MC} and consider in detail the case where there are two energy donors and one acceptor.

The importance of understanding the efficient and fast energy transfer processes like the ones involved in Photosynthesis~\cite{psbook} cannot be understated and over the past few years, there is a growing belief that quantum coherence and entanglement may be enabling resources for these processes~\cite{entnglmnt}. Direct evidence  using two dimensional fourier transform electronic spectroscopy~\cite{2dspec1,2dspec2,2dspec3,2dspec1} methods reveal remarkably long-lived quantum coherences in Fenna-Matthews-Olson (FMO) protein complex~\cite{engel_evidence_2007}, within photosynthetic structures. The coherences that are observed are both vibrionic and electronic in nature and as such there is indirect evidence that these coherences may have a role to play in the photosynthetic processes~\cite{Renger:PhysChemChemPhys:2013}.  Given the intricacies of the photosynthetic complex there is also some evidence that even the immediate environment of the chromophores that directly participate in the energy transfer process have been engineered by nature to enhance the coherence assisted transport instead of being detrimental to it as is the norm~\cite{env1,env2,env3,env4,env5}.  

In this Paper we explore in detail a simple model in which energy is transferred via resonance transfer from two energy donors to a single acceptor. We assume that there can be quantum mechanical coherences between the two donor molecules and we also assume that the entire system is in contact with a rather simplified and mesoscopic (low dimensional) ``environment''. We look for the signatures of  coherence in the energy transfer rate between the donors and the acceptor as well as for signs that under specific conditions the effect of the environment is to influence the energy transfer process positively by enhancing its efficiency and rate.  We derive the expression for the rate of energy transfer between independent donors which are coherently excited into the single excitation section and an acceptor. Multiple donors being coherently and simultaneously excited is a very plausible scenario in the context of photosynthetic processes in light of the fact that a single photon is typically 'bigger' than the photosynthetic complex itself. The donors are not typically independent of each other in the photosynthetic complex since they are closely packed together. However in what follows we assume for simplicity that the energy donors are not coupled to each other. Note that the development in the following can be extended to a  system of strongly coupled donors by considering the normal modes of the coupled system rather than the individual levels of the independent constituents.  A comparison of the analytical rate expressions for a mesoscopic environment indicate that the enhancement of the rate due to the coherent donors may be measurable at short times. The initial enhancement to the energy transfer rate is expressed as a measurable spectral overlap integral which would be detectable at high frequencies in a pump probe experiment.

This Paper is organised as follows: In the next section we briefly recap FRET and its extensions to the multiple donor case with reference to the model we are considering. In Section~\ref{mesoscopic}, we look at the mesoscopic environment and its effects on the energy transfer rate. We also show a way of computing the effect of the environment utilising Wigner functions. Our conclusions are in Section~\ref{conclusion}. 

\section{FRET with multiple donors and coherence \label{sec2}}

The incoherent energy hopping mechanisms for energy transfer proposed by F{\"o}rster~\cite{fster1,fster2,fster3} and Dexter~\cite{dexter} was generalised to account for short time nonequilibrium kinetics as well as for multiple donor and acceptor case  by  Jang et. al~\cite{noneq,MC}. F\"{o}ster's approach, as outlined in~\cite{coh3,coupling} applies to two chromophores, one being the donor and the other the energy acceptor. Each chromophore has two valance electrons with spins denoted by $\alpha$ and $\beta$. Let  $\phi_{\alpha}(\psi_{\alpha})$ and $\phi_{\beta}(\psi_{\beta})$ denote the spin orbitals in the Highest Occupied Molecular Orbitals (HOMO) of the donor(acceptor) chromophore respectively. The corresponding Lowest Unoccupied Molecular Orbitals (LUMO) are labelled as $\phi_{\beta(\alpha)}^{*}(\psi_{\beta(\alpha)}^{*})$. It is further assumed that the orbitals localized on the same chromophore are orthonormal, though inter chromophore orbital overlaps are allowed. The states with complete localization of excitation in the donor (denoted by $|D\rangle$) and the acceptor (denoted by $|A\rangle$) are 
\begin{eqnarray*}
| D\rangle & = &  \gamma_{1}(|\phi_{\alpha}^{*}\psi_{\alpha}\phi_{\beta}\psi_{\beta}| + |\phi_{\alpha}\psi_{\alpha}\phi_{\beta}^{*}\psi_{\beta}|)\\
| A\rangle & = &  \gamma_{2}(|\phi_{\alpha}\psi_{\alpha}^{*}\phi_{\beta}\psi_{\beta}| + |\phi_{\alpha}\psi_{\alpha}\phi_{\beta}\psi_{\beta}^{*}|), 
\end{eqnarray*}
where $\gamma_{1}$ and $\gamma_{2}$ are the normalization factors. The states are Slater determinants since there can be orbital overlap between the chromophores making all the electrons indistinguishable from one another. The rate of energy transfer is obtained starting from the matrix element describing the transition from the $|D\rangle$ state in which the excitation is localized in the donor chromophore to the $|A\rangle$ state in which it is localized in the acceptor. We can write this matrix element as
\begin{equation} 
	\label{eq:vda}
	V_{DA} = \langle D| \hat{H}| A\rangle \simeq 2 (\phi^{*}\phi|\psi\psi^{*}) - 2 (\phi^{*}\psi^{*}|\psi\phi) + O(\langle\psi|\phi\rangle^2) + O(\langle\psi|\phi\rangle^3)+\dots, 
	\end{equation}
where
\[ (ab| cd) \equiv \langle \Psi_{a}(i)\Psi_{c}(j)| r_{ij}^{-1}| \Psi_{b}(i)\Psi_{d}(j)\rangle. \]
The relative strengths of the various terms in Eq.~(\ref{eq:vda}) depends on the separation $r_{ij}$ between the chromophores. At short range (3-6  {\AA}), both orbital overlap effects and coulomb effects are relevant. In the intermediate range (6-20  {\AA}), the orbital overlaps can be ignored and only the electromagnetic interaction, which is typically dipole-dipole, is relevant. At long range one has to include the modifications to the dipole-dipole interaction adding retardation effects etc that arise from real photons being emitted and re-absorbed~\cite{Andrews:JChemPhys:1992}. F\"{o}ster's theory was originally developed for the intermediate regime where the orbital overlaps are small and so only the first term in Eq.~(\ref{eq:vda}) contributes. This term is a two electron integral that describes the de-excitation of the donor and the excitation of the acceptor that happens simultaneously. The integral can be thought of as the coulomb interaction between the two transition densities. These transition densities, in turn, can be approximated - through a multipole expansion - by dipoles and the interaction reduces to a dipole-dipole coupling with a characteristic $1/r^{6}$ dependence for the transition rate as a function of the distance between the chromophores. For completeness, it  may be noted that Dexter's theory~\cite{dexter} holds good for the transition rate when the distances between the chromophores is very short. 

In this Paper we focus on the case where there is a single excitation present in a system with multiple energy donors and a single acceptor. The separation between the donors is assumed to be small so that orbital overlap effects are relevant and in particular there can be quantum coherences between the donor chromophores. In the discussions that follow, the energy acceptor is well separated from the donor chromophores so that orbital overlap effects between the donors and the acceptor are not relevant. To keep the focus on the effects of coherence and on the influence of the structured environment we resist the temptation to give the donors and acceptors realistic and complicated level structures~\cite{realchrome1,realchrome2} and treat them as two level quantum systems (qubits) as is done in much of the existing literature~\cite{MC,coh1,coh2,coh3,engel_evidence_2007,entnglmnt,noneq,review}. 

\subsection{The model and energy transfer rates}

We are looking only at the single excitation sector of the system with two energy donors and one acceptor. Let $|g\rangle$ denote the ground state of all three chromophores. The states of interest to us are $|D_{1} \rangle = a_{1}^{\dagger}|g\rangle$, $|D_{2} \rangle = a_{2}^{\dagger}|g\rangle$, and $|A \rangle = a_{A}^{\dagger}|g\rangle$. The chromophores are assumed to be sitting in a noisy environment or bath. The Hamiltonian for the system is~\cite{noneq}
\[ H  = H_{0} + V, \]
where
\begin{equation}
\label{eq:h0}
H_{0} = \epsilon_{A}a_{A}^{\dagger} a_{A}^{\vphantom{\dagger}} +  \epsilon_{1}a_{1}^{\dagger} a_{1}^{\vphantom{\dagger}} + \epsilon_{2}a_{2}^{\dagger} a_{2}^{\vphantom{\dagger}} + H_{eb} + H_{b}.
\end{equation}
Here $\epsilon_{1(2)}$ and $\epsilon_{A}$ are the energies of the excited states of donor(s) and the acceptor respectively,  while $H_{eb}$ is the excitation bath coupling of the form
\begin{equation}
\label{eq:Heb}
H_{eb} = B_{1}a_{1}^{\dagger} a_{1}^{\vphantom{\dagger}} +  B_{2}a_{2}^{\dagger} a_{2}^{\vphantom{\dagger}} +B_{A}a_{A}^{\dagger} a_{A}^{\vphantom{\dagger}}, 
\end{equation}
Where $B_{1}$, $B_{2}$ and $B_{A}$ are bath operators that will be specified later on. $H_{b}$ denotes the Hamiltonian for the bath. We assume that there are no bath modes that are coupled to more than one of the three chromophores at a time. This means that energy transfer from the donors to the acceptor cannot be mediated by the bath. This assumption of having no common bath modes implies that we can view the bath as made of three disconnected pieces so that
\[ H_{b} = H_{b1} + H_{b2} + H_{bA}, \]
with the three terms in the sum representing the Hamiltonians for the parts of the bath coupled to $D_{1}$, $D_{2}$ and $A$ respectively. We also have
\begin{equation}
	\label{eq:comm1} [H_{bj}, H_{bk}] = [H_{bj}, B_{k}] = [B_{j}, B_{k}]  = 0 \quad {\rm for} \quad {j \neq k}, \quad j,k=1,2,A,
\end{equation}
in addition to the standard commutation relations, $[H_{bj}, H_{bj}] = [B_{j},B_{j}]=0$.

The resonant interaction between the donors and the acceptor is the perturbation $V$, 
\begin{equation}
	\label{eq:perturb} 
	V = J_{1} a_{1}^{\dagger}a_{A}^{\vphantom{dagger}} +  J_{2} a_{2}^{\dagger}a_{A}^{\vphantom{dagger}} + {\mbox{H. C.}}
\end{equation}
In treating the interaction as a perturbation we are assuming that the interaction strengths $J_{i}$ between the donors and the acceptor is small owing to the assumption of relatively large separation between the two. The transition probability for the excitation in the donors to move to the acceptor is given by
\[ p_{A}(t) = {\rm tr}_{b} \langle A_{I}| \rho_{I}(t) | A_{I} \rangle, \]
with the index $I$ indicating that the above expression is in the interaction picture. We consider an initial state for the system and the bath of the form
\begin{equation}
	\label{eq:initialstate} 
	\rho_{0} = \frac{1}{Z_{b}} e^{-\beta H_{b}} |\psi \rangle \langle \psi|, 
\end{equation}
where
\begin{equation}
	\label{eq:IniState} 
	|\psi \rangle = \sqrt{p}|D_{1} \rangle + e^{-i\phi}\sqrt{1-p}|D_{2} \rangle. 
\end{equation}
In other words, initially the single excitation is delocalized between the two donors with the donors in an superposed state.

Following closely the discussions in \cite{MC,noneq}, we obtain the following expression for the transition probability (See Appendix~\ref{appendixA} for more details), 
\begin{eqnarray}
	\label{eq:transitionprob}
p_{A}(t) & = & p \frac{ J_{1}^{2}}{ Z_{b}}  \int_{0}^{t}  dt' \int_{0}^{t} dt'' \, e^{i\epsilon_{A}(t'-t'')} e^{-i\epsilon_{1}(t'-t'')} \nonumber \\
&& \qquad \qquad \times \;  {\rm tr}_{b} \big[ e^{i(B_{A}+H_{b})(t'-t'')}e^{-i(B_{1}+H_{b})t'}e^{-\beta H_{b}}e^{i(B_{1}+H_{b})t''}\big]\nonumber \\
&&  +  (1-p) \frac{J_{2}^{2}}{ Z_{b}} \int_{0}^{t}  dt' \int_{0}^{t}  dt'' \, e^{i\epsilon_{A}(t'-t'')}  e^{-i\epsilon_{2}(t'-t'')} \nonumber \\
&& \qquad \qquad \times \; {\rm tr}_{b} \big[ e^{i(B_{A}+H_{b})(t'-t'')} e^{-i(B_{2}+H_{b})t'}e^{-\beta H_{b}}e^{i(B_{2}+H_{b})t''}\big] \nonumber \\
&& +  e^{i\phi} \sqrt{p} \sqrt{1-p} \frac{J_{1}J_{2}}{ Z_{b}}  \int_{0}^{t} dt'  \int_{0}^{t}  dt'' \,  e^{i\epsilon_{A}(t'-t'')} e^{-i\epsilon_{1}t'}e^{i\epsilon_{2}t''} \nonumber \\
&& \qquad \qquad \times \; {\rm tr}_{b} \big[ e^{i(B_{A}+H_{b})(t'-t'')} e^{-i(B_{1}+H_{b})t'} e^{-\beta H_{b}} e^{i(B_{2}+H_{b})t''} \big] \nonumber \\
&& + e^{-i\phi} \sqrt{p} \sqrt{1-p}  \frac{J_{2}J_{1}}{ Z_{b}}  \int_{0}^{t}  dt'  \int_{0}^{t}  dt'' \, e^{i\epsilon_{A}(t'-t'')} e^{-i\epsilon_{2}t'}e^{i\epsilon_{1}t''} \nonumber \\
&& \qquad \qquad \times \;  {\rm tr}_{b} \big[ e^{i(B_{A}+H_{b})(t'-t'')} e^{-i(B_{2}+H_{b})t'}e^{-\beta H_{b}}e^{i(B_{1}+H_{b})t''} \big]. 
\end{eqnarray}
The energy transfer rate is the derivative of the transition probability:
\begin{eqnarray}
	\label{eq:transrate}
	k(t) & = &  \dfrac{d}{dt}p_{A}(t) \nonumber \\
	& = &  2\:{\rm Re}\:\bigg\{ p\frac{J_{1}^{2}}{Z_{b1}Z_{bA}} \int^{t}_{0}dt' e^{i(\epsilon_{A}-\epsilon_{1})(t-t')} {\rm tr}_{b_{1}}\big[e^{i (B_{1}+H_{b1})t'} e^{i H_{b1}(t-t')} e^{-i (B_{1}+H_{b1})t} e^{-\beta H_{b1}} \big] \nonumber  \\
	&& \qquad \qquad \times  {\rm tr}_{b_{A}}\big[ e^{i (B_{A}+H_{bA})(t-t')} e^{-i H_{bA}(t-t')} e^{-\beta H_{bA}}\big] \nonumber  \\
	&&  + (1-p)  \dfrac{J_{2}^{2}}{Z_{b2}Z_{bA}} \int^{t}_{0}dt' e^{i(\epsilon_{A}-\epsilon_{2})(t-t')} {\rm tr}_{b_{2}}\big[  e^{i (B_{2}+H_{b2})t' e^{i H_{b2}(t-t')} e^{-i (B_{2}+H_{b2})t} e^{-\beta H_{b2}}}\big] \nonumber \\
	&& \qquad \qquad \times  {\rm tr}_{b_{A}}\big[ e^{i (B_{A}+H_{bA})(t-t')} e^{-i H_{bA}(t-t')} e^{-\beta H_{bA}}\big] \nonumber \\
	&&  + e^{i\phi} \sqrt{p}\sqrt{1-p}\dfrac{J_{1}J_{2}}{Z_{b1}Z_{b2}Z_{bA}} \int^{t}_{0}dt' e^{i(\epsilon_{A}-\epsilon_{1})t} e^{-i(\epsilon_{A}-\epsilon_{2})t'} {\rm tr}_{b_{1}}\big[ e^{i H_{b1}t} e^{-i (B_{1}+H_{b1})t} e^{-\beta H_{b1}}\big] \nonumber \\
	&& \qquad \qquad \times {\rm tr}_{b_{2}} \big[ e^{i (B_{2}+H_{b2})t'} e^{-i H_{b2}t'} e^{-\beta H_{b2}}\big]   {\rm tr}_{b_{A}}\big[ e^{i (B_{A}+H_{bA})(t-t')} e^{-i H_{bA}(t-t')} e^{-\beta H_{bA}}\big] \nonumber \\
	&&  +  e^{-i\phi} \sqrt{p}\sqrt{1-p} \dfrac{J_{2}J_{1}}{Z_{b2}Z_{b1}Z_{bA}} \int^{t}_{0}dt' e^{i(\epsilon_{A}-\epsilon_{2})t} e^{-i(\epsilon_{A}-\epsilon_{1})t'} {\rm tr}_{b_{2}}\big[ e^{i H_{b2}t} e^{-i (B_{2}+H_{b2})t} e^{-\beta H_{b2}}\big] \nonumber \\
	&& \qquad \qquad \times {\rm tr}_{b_{1}} \big[ e^{i (B_{1}+H_{b1})t'} e^{-i H_{b1}t'} e^{-\beta H_{b1}}\big]  {\rm tr}_{b_{A}}\big[ e^{i (B_{A}+H_{bA})(t-t')} e^{-i H_{bA}(t-t')} e^{-\beta H_{bA}}\big]\bigg\}.
\end{eqnarray}
For separating out the traces over the three sets of bath modes, we have used the commutation relations in Eq.~(\ref{eq:comm1}). 

Since there is only one acceptor, we are able to factorize out the contribution to the rate expression from the acceptor within the integral as
\begin{equation}
	\label{eq:FT} 
	\frac{1}{Z_{bA}} e^{i\epsilon_{A}(t-t')} {\rm tr}_{b_{A}}\big[ e^{i (B_{A}+H_{bA})(t-t')} e^{-i H_{bA}(t-t')} e^{-\beta H_{bA}}\big]  = \frac{1}{\sqrt{2\pi}|\vec{\mu}_{A} \cdot \hat{e}|^{2}}  \int_{-\infty}^{\infty} d\omega \, e^{i \omega (t-t')} I_{A}(\omega), 
\end{equation}
where $\vec{\mu}_{A}$ is the transition dipole moment of the acceptor, $\hat{e}$ is a reference axis taken the polarization vector of the incident radiation if the acceptor is irradiated to find its absorption profile, and 
\[ I_{A}(\omega) \equiv |\vec{\mu}_{A} \cdot \hat{e}|^{2} \frac{1}{Z_{bA}\sqrt{2 \pi}} \int_{-\infty}^{\infty} du\, e^{-i\omega u} e^{i\epsilon_{A}u}{\rm tr}_{b_{A}}\big[ e^{i (B_{A}+H_{bA})u} e^{-i H_{bA}u} e^{-\beta H_{bA}}\big],\]
is the absorption profile of $A$. In terms of the absorption profile, we can re-write the expression for the energy transfer rate as 
\begin{eqnarray}
	\label{eq:rate2}
	k(t) & = &  \frac{2}{\sqrt{2\pi}|\vec{\mu}_{A} \cdot \hat{e}|^{2}}  \int_{-\infty}^{\infty} d\omega \, I_{A}(\omega) \, {\rm Re} \int_{0}^{t} dt' \, e^{i\omega(t-t')} \, \nonumber \\
	&& \; \times \bigg\{ p \frac{J_{1}^{2}}{Z_{b1}} e^{-i\epsilon_{1} (t-t')}  {\rm tr}_{b_{1}}\big[ e^{i H_{b1}(t-t')} e^{-i (B_{1}+H_{b1})t} e^{-\beta H_{b1}} e^{i (B_{1}+H_{b1})t'}\big] \nonumber \\
	&&  \;\; + \; (1-p)\frac{J_{2}^{2}}{Z_{b2}} e^{-i\epsilon_{2}(t-t')} {\rm tr}_{b_{2}}\big[ e^{i H_{b2}(t-t')} e^{-i (B_{2}+H_{b2})t} e^{-\beta H_{b2}} e^{i (B_{2}+H_{b2})t'}\big] \nonumber \\
	&&  \;\; + \;  e^{i\phi} \sqrt{p}\sqrt{1-p} \frac{J_{1}J_{2}}{Z_{b1}Z_{b2}} e^{-i\epsilon_{1}t+i\epsilon_{2} t'} {\rm tr}_{b_{1}}\big[ e^{i H_{b1}t} e^{-i (B_{1}+H_{b1})t} e^{-\beta H_{b1}} \big] \nonumber \\
	&& \qquad \qquad \qquad \qquad \times \; {\rm tr}_{b_{2}} \big[ e^{i (B_{2}+H_{b2})t'} e^{-i H_{b2}t'} e^{-\beta H_{b2}}\big]   \nonumber \\
	&&  \;\; + \;   e^{-i\phi} \sqrt{p}\sqrt{1-p}  \frac{J_{1}J_{2}}{Z_{b1}Z_{b2}} e^{-i\epsilon_{2}t+i\epsilon_{1} t'} {\rm tr}_{b_{2}}\big[ e^{i H_{b2}t} e^{-i (B_{2}+H_{b2})t} e^{-\beta H_{b2}} \big] \nonumber \\
	&& \qquad \qquad \qquad \qquad \times \; {\rm tr}_{b_{1}} \big[ e^{i (B_{1}+H_{b1})t'} e^{-i H_{b1}t'} e^{-\beta H_{b1}}\big]\bigg\}. \;\;\;
\end{eqnarray}

Our objective is to connect the expression for the energy transfer rate in Eq.~(\ref{eq:rate2}) to the time dependent emission profile for a coherently excited initial state of the two donors within the single excitation manifold. With this in mind, we start with the initial state for the donors and their environment given in Eqs~(\ref{eq:initialstate}) and (\ref{eq:IniState}) and the environment Hamiltonian redefined in a reduced manner excluding the environment of the acceptor as 
\[ H_{bD} \equiv H_{b1} + H_{b2}. \]
The stimulated emission profile is obtained by placing the coherently excited pair of donors in an electromagnetic field of frequency $\nu$ and polarization $\hat{e}$. Assuming unit field strength and using the rotating wave approximation, the Hamiltonian governing the dynamics of the stimulated emission process is
\[ H(t) = H_{D} + V(t), \] 
where
\[ H_{D} =  \epsilon_{1}a_{1}^{\dagger} a_{1}^{\vphantom{\dagger}} + \epsilon_{2}a_{2}^{\dagger} a_{2}^{\vphantom{\dagger}} +  B_{1}a_{1}^{\dagger} a_{1}^{\vphantom{\dagger}} +  B_{2}a_{2}^{\dagger} a_{2}^{\vphantom{\dagger}} +H_{bD}, \]
and
\[V(t) =  \vert \vec{\mu}_{1} \cdot \hat{e}\vert  (e^{-i\nu t}a_{1}^{\dagger}+ e^{i\nu t} a_{1} ) + \vert \vec{\mu}_{2} \cdot \hat{e}\vert ( e^{-i\nu t}a_{2}^{\dagger} + e^{i\nu t} a_{2} ).\]
Using the interaction picture we can write down the probability that a stimulated emission of a photon occurs and the two donors come to their respective ground state, $|0 \rangle$ as
\[ p_{\nu}(t) = {\rm tr}_{bD} \big[ \langle 0_{I} | e^{-i \int_{0}^{t} V_{I}(t') dt'} \rho_{0I}^{Db} e^{i \int_{0}^{t} V_{I}(t'') dt''}| 0_{I} \rangle  \big], \]
with $\vec{\mu}_{j}$, $j=1,2$ denoting the induced molecular dipole moments of each of the two donors. Proceeding along the same lines as described in Appendix~\ref{appendixA}, in the weak field limit, where we expand the exponential above to first order in $|\vec{\mu}_{j} \cdot \hat{e}|$, we obtain the time dependent stimulated emission profile which is the time derivative of the emission probability as:
\begin{eqnarray}
	\label{eq:emission1}
	E_{\psi}(\nu, t) & = & \frac{d\;}{dt} p_{\nu}(t) \nonumber \\
	& = & 2\:{\rm Re}\:\bigg\{ p \frac{ \vert \vec{\mu}_{1} \cdot \hat{e}\vert^{2}}{Z_{b1}} \int_{0}^{t}dt' e^{i\nu(t-t')} e^{-i\epsilon_{1}(t-t')} \nonumber \\
	&& \qquad \qquad \qquad \times \; {\rm tr}_{b_{1}} \big[ e^{iH_{b1}(t-t')}e^{-i(B_{1}+H_{b1})t}e^{-\beta H_{b1}}e^{i(B_{1}+H_{b1})t'} \big] \nonumber \\ 
	&& +  (1-p)\frac{ \vert \vec{\mu}_{2} \cdot \hat{e}\vert^{2}}{Z_{b2}}  \!\! \int_{0}^{t} \! \!  dt' e^{i\nu(t-t')} e^{-i\epsilon_{2}(t-t')} \nonumber \\
	&& \qquad\qquad\qquad \times \; {\rm tr}_{b_{2}} \big[  e^{iH_{b2}(t-t')} e^{-i(B_{2}+H_{b2})t} e^{-\beta H_{b2}}e^{i(B_{2}+H_{b2})t' } \big] \nonumber \\
	&& + \sqrt{p}\sqrt{1-p}e^{i\phi} \frac{ \vert \vec{\mu}_{1} \cdot\hat{e}\vert\vert \vec{\mu}_{2} \cdot\hat{e}\vert}{Z_{b1}Z_{b2}} \! \int_{0}^{t}dt' e^{i\nu(t-t')}e^{-i\epsilon_{1}t}e^{i\epsilon_{2}t'} \nonumber \\
	&& \qquad \qquad \qquad \times \; {\rm tr}_{b1} \big[  e^{iH_{b1}t}e^{-i(B_{1}+H_{b1})t}e^{-\beta H_{b1}}\big] {\rm tr}_{b2} \big[ e^{i(B_{2}+H_{b2})t'}  e^{-iH_{b2}t'}e^{-\beta H_{b2}}\big] \nonumber \\
	&& + \sqrt{p}\sqrt{1-p}e^{-i\phi}  \frac{ \vert \vec{\mu}_{2} \cdot\hat{e}\vert\vert \vec{\mu}_{1} \cdot\hat{e}\vert}{Z_{bD}} \! \int_{0}^{t}dt' e^{i\nu(t-t')}e^{-i\epsilon_{2}t}e^{i\epsilon_{1}t'} \nonumber \\
	&& \qquad \qquad \qquad \times \; {\rm tr}_{b2} \big[  e^{iH_{b2}t}e^{-i(B_{2}+H_{b2})t}e^{-\beta H_{b2}}\big] {\rm tr}_{b1} \big[ e^{i(B_{1}+H_{b1})t'}  e^{-iH_{b1}t'}e^{-\beta H_{b1}}\big]\bigg\}.
\end{eqnarray}

\subsection{Identical donors}

Let us now specialize to the case where we have identical donors with $\epsilon_{1} = \epsilon_{2} = \epsilon_{D}$ having identical couplings $J_{1} = J_{2} = J$ to the acceptor chromophore and symmetrically placed with respect to the acceptor so that $ | \vec{\mu}_{1} \cdot\hat{e}| = | \vec{\mu}_{2} \cdot \hat{e} | = |\vec{\mu}_{D} \cdot \hat{e} |$. Such a system can potentially be realised in the lab by arranging suitable donor and acceptor chromophores on a molecular scaffold like a DNA strand. This will allow for a bottom up approach to the larger problem of understanding the role of quantum coherence in biologically relevant processes by starting with simpler non-biological systems with fewer chromophores involved. For the identical donors we can write the rate equation (\ref{eq:rate2}) as
\begin{equation}
	\label{eq:emission2}
	k(t)  =   \frac{2J^{2}}{\sqrt{2 \pi} |\vec{\mu}_{A} \cdot \hat{e}|^{2} |\vec{\mu}_{D} \cdot \hat{e}|^{2} }  \int_{-\infty}^{\infty} d\omega \, I_{A}(\omega) E_{\psi}(\omega, t).
\end{equation}
The emission profile $E_{\psi}$ for identical donors can be written as
\[ E_{\psi} (\nu, t)= E_{D} (\nu, t)+ E_{\rm coh} (\nu, t), \]
where
\begin{eqnarray}
	\label{eq:ed} 
	E_{D}(\nu, t) & = & 2 |\vec{\mu}_{D} \cdot \hat{e}|^{2}  {\rm Re} \int_{0}^{t} dt' \, e^{i(\nu -\epsilon_{D}) (t-t')} \nonumber \\
	&& \quad \times \; \bigg\{ \frac{p}{Z_{b1}} {\rm tr}_{b_{1}} \big[ e^{iH_{b1}(t-t')}e^{-i(B_{1}+H_{b1})t}e^{-\beta H_{b1}}e^{i(B_{1}+H_{b1})t'} \big] \nonumber \\
	&& \qquad + \; \frac{1-p}{Z_{b2}}  {\rm tr}_{b_{2}} \big[  e^{iH_{b2}(t-t')} e^{-i(B_{2}+H_{b2})t} e^{-\beta H_{b2}}e^{i(B_{2}+H_{b2})t' } \big]  \bigg\},
\end{eqnarray}
and
\begin{eqnarray}
	\label{eq:cog} 
	E_{\rm coh}(\nu, t) & = & 2 \sqrt{p} \sqrt{1-p} \frac{|\vec{\mu}_{D} \cdot \hat{e}|^{2} }{Z_{b1}Z_{b2}} {\rm Re} \int_{0}^{t} dt' \, e^{i(\nu -\epsilon_{D}) (t-t')} \nonumber \\
	&& \quad \times \; \Big\{ e^{i\phi} \, {\rm tr}_{b1} \big[  e^{iH_{b1}t}e^{-i(B_{1}+H_{b1})t}e^{-\beta H_{b1}}\big] {\rm tr}_{b2} \big[ e^{i(B_{2}+H_{b2})t'}  e^{-iH_{b2}t'}e^{-\beta H_{b2}}\big]  \nonumber \\
	&& \qquad + \; e^{-i\phi} \, {\rm tr}_{b2} \big[  e^{iH_{b2}t}e^{-i(B_{2}+H_{b2})t}e^{-\beta H_{b2}}\big] {\rm tr}_{b1} \big[ e^{i(B_{1}+H_{b1})t'}  e^{-iH_{b1}t'}e^{-\beta H_{b1}}\big] \Big\}.
\end{eqnarray}

Notice that if we either set $p=1$ (or $p=0$), or assume that the baths associated with each of the donors is identical making the two donors identical in all respects $E_{D}(\nu, t)$ reduces to the emission profile of a single donor as used in F\"{o}ster's theory~ \cite{noneq}. For $p=0$ or $p=1$, $E_{\rm coh}(\nu, t)=0$ while for identical baths and $p=1/2$ we have,
\begin{eqnarray}
	 E_{\rm coh}(\nu, t)  & = &  2 \cos \phi \frac{|\vec{\mu}_{D} \cdot \hat{e}|^{2} }{Z_{b}} {\rm Re} \int_{0}^{t} dt' \, e^{i(\nu -\epsilon_{D}) (t-t')} \nonumber \\
	 && \qquad \qquad \times \;  {\rm tr}_{b'} \big[  e^{iH_{b'}t}e^{-i(B+H_{b'})t}e^{-\beta H_{b'}}\big] {\rm tr}_{b'} \big[ e^{i(B+H_{b'})t'}  e^{-iH_{b'}t'}e^{-\beta H_{b'}}\big], 
\end{eqnarray}
where the subscript $b'$ denotes the bath attached to one of the donors and the subscript $b$ denotes the entire bath with $B_{1} = B_{2} = B$. The form of $E_{\rm coh}$ suggests that in Eq.~(\ref{eq:emission2}), it acts as an interference term modulating the energy transfer rate depending on the relative phase $\phi$ of the initial superposition in  Eq.~(\ref{eq:IniState}). We see that the electronic coherence upon photon absorption between the donors can either enhance or suppress the energy transfer rate relative to that of a single donor within the single excitation manifold.

\section{Mesoscopic environment models \label{mesoscopic}}

We consider two types of low dimensional quantum systems as the bath modes coupled to each of the two donors and the acceptor in the following. In the first case we assume that the bath attached to each is a single Harmonic oscillator and in the second case we assume that the bath is a collection of $N$ qubits where $N$ is relatively small. As mentioned earlier, a mesoscopic environment allows us to numerically integrate the Schr\"{o}dinger equation for the entire system including the bath and compare with the analytic results in the previous section. More importantly relative simplicity of the bath lets us clearly see and separate out the bath effects in the dynamics from the effect of the coherence between the donors. 

The numerical computations are done in arbitrary units assuming $\hbar=1$. However to put the results we obtain in context it is necessary to make the connection with the energy, time and distance scales relevant to some of the systems that have been studied in detail previously. Following up on~\cite{MC}, in~\cite{Jang:JPhysChemB:2007}, the rate predicted by the Multi-chromophoric generalisation of FRET for energy transfer between the B800 unit to the B850 unit in the light harvesting complex 2 of purple bacteria is computed. As a prototype for providing the context for our results which are oriented towards qualitative understanding of the role of coherence (and hence in arbitrary units) we use the system in~\cite{Jang:JPhysChemB:2007}. The excitation energy of the B850 unit, which has the role of acceptor in the system studied in~\cite{Jang:JPhysChemB:2007}, is around $2 \times 10^{-19}$ Joules. In the numerical computations that follow we have taken the excitation energies $\epsilon_{1} = \epsilon_{2} = \epsilon_{A} = 0.1$ in arbitrary units. Inverse of our excitation energy (in units of $\hbar=1$) is then equal to 10/$2\pi$ time units. Therefore, in relation to the system considered in~\cite{Jang:JPhysChemB:2007}, one unit of time in the numerical examples below corresponds to around 2 femto-seconds. An analogous scaling for the basic time unit can be constructed for other realistic systems as well like the one discussed previously with chromophores attached to DNA structures knowing the excitation energies of the chromophores. The coupling between the chromophores in~\cite{Jang:JPhysChemB:2007} is characterised by an interaction energy of approximately $2 \times 10^{-20}$ Joules computed assuming a dipole moment of 10 Debyes and an intra-chromophore separation in vacuum of around 20 {\AA}. Accordingly we have taken the perturbative coupling between the donors and the acceptor with a characteristic interaction energy of 0.01 in our arbitrary units.

\subsection{Harmonic oscillator bath}

Attached to each of the chromophores is single harmonic oscillator taken to be the bath. The bath Hamiltonian is:
\begin{equation}
	\label{eq:bath1H}
	H_{b} = \sum_{s}w_{s}\bigg( b_{s}^{\dagger}b_{s}^{\vphantom{dagger}} + \frac{1}{2} \bigg), \qquad s = 1,2,A,
\end{equation}
where $b_{s}$ ($b_{s}^{\dagger}$) is the annihilation (creation) operator for the harmonic oscillator of frequency $w_{s}$ coupled to chromophore $s$. The system bath coupling is assumed to be linear and of the form,
\begin{equation}
	\label{eq:coupling1H}
	H_{eb} = \sum_{s} g_{s}(b_{s}^{\dagger} + b_{s}^{\vphantom{dagger}})a_{s}^{\dagger}a_{s}^{\vphantom{dagger}} . 
\end{equation}
The system-bath coupling chosen along the lines of the dispersive coupling in cavity opto-mechanics~\cite{kippenberg_cavity_2008,kippenberg_cavity_2007,jacobs_quantum-nondemolition_1994,jacobs_quantum_1999} is such that the number of excitations shared between the donors and acceptors is conserved. Time evolution of the entire system consisting of the chromophores and their respective baths is done numerically taking only the first few levels of each harmonic oscillator into consideration. Starting from the initial state in Eq.~(\ref{eq:initialstate}) we numerically integrate the Schr\"{o}dinger equation for the system and compute the population in $|A\rangle$ as a function of time. Time derivative of the population gives us the energy transfer rate.   

To evaluate the analytic expression we have for the energy transfer rate in Eq.~(\ref{eq:transrate}) we have to find  expectation values of products of exponentials of the form $e^{\pm i(B+H_{b}) \tau}$ and $e^{\pm i H_{b} \tau}$ with respect to the canonical state $e^{-\beta H_{b}}/Z_{b}$ of the bath. We can compute these expectation values as follows. As an example consider the term of the form ${\rm tr}_{b}[e^{i (B+H_{b})t'} e^{i H_{b}(t-t')} e^{-i (B+H_{b})t} e^{-\beta H_{b}} ] $, appearing in Eq.~(\ref{eq:transrate}). For short times $t$, we can apply the Baker-Campbell-Hausdroff formula~~\cite{baker} for $Q = e^{U}e^{V}e^{W}$ as
\[ \ln Q = U + V+ W + \dfrac{1}{2}( [U,W] + [U,V] + [V,W] ) + \ldots, \]
and write the following expression correct to second order in $t$, $t'$ and $t-t'$:
\[ e^{i (B+H_{b})t'}e^{i H_{b}(t-t')}e^{-i (B+H_{b})t} \simeq e^{ -\frac{ig}{ \sqrt{2 w}}  (t-t') [  2 w \hat{x} + w (t+t')  \hat{p} ]} \equiv \hat{O}_{2}(\hat{x}, \hat{p}|t,t'), \]
where $\hat{x}$ and $\hat{p}$ are the position and momentum operators of the harmonic oscillators. Here we have used the canonical commutation relations of the bath operators and also assumed that the mass of the harmonic oscillators are all equal to unity. The expectation value $\langle \hat{O}_{2}(\hat{x}, \hat{p}|t,t') \rangle $ can be computed using the Wigner function $W(x,p)$ of the state $\rho$ of the bath oscillator as~\cite{wigner},
\begin{equation}
	\label{eq:wigavg}
	 {\rm tr}[\hat{O}_{2}(\hat{x}, \hat{p}|t,t') \rho] = \int dx \int dp \, O_{2}(x, p|t,t') W(x,p). 
\end{equation}
The Wigner function for a single mode thermal state is
\[ W(x,p) = \frac{e^{-\frac{1}{2} \eta^{T} V^{-1} \eta}}{2 \pi \sqrt{\det (V)}}, \qquad \eta = (x,p)^{T}, \]
where $V$ is the variance matrix and the superscript $T$ denotes the transpose operation. Using
\[ V = \left( \begin{array}{cc} \frac{1}{2} \coth \frac{\beta w}{2} & 0 \\ 0 &  \frac{1}{2} \coth \frac{\beta w}{2} \end{array} \right),  \qquad \beta = \frac{1}{T},\]
for a normalised thermal state, $e^{-\beta H_{b}}/Z_{b}$, we get 
\[ W(x,p) = \frac{1}{\pi} \tanh \frac{\beta w}{2} e^{-\tanh \frac{\beta w}{2} (x^{2} + p^{2})}. \]
Using Eq.~(\ref{eq:wigavg}) we obtain,
\begin{equation}
	\label{eq:wig1} 
	\langle \hat{O}_{2}(t,t') \rangle = e^{-\frac{g w}{8}(t-t')^{2} [(t+t')^{2} + 4] \coth \frac{\beta w}{2} }. 
\end{equation}
In a similar manner we find
\begin{equation}
	\label{eq:wig2}	
	\frac{1}{Z_{b}} {\rm tr}_{b}\big[e^{i (B+H_{b})(t-t')} e^{-i H_{b}(t-t')} e^{-\beta H_{b}} \big]  \simeq  \langle \hat{O}_{1}(t,t') \rangle  =  e^{-\frac{g w}{8}(t-t')^{2} [(t-t')^{2} + 4] \coth \frac{\beta w}{2}} 
\end{equation}
and
\begin{equation}
	\label{eq:wig3}	
	\frac{1}{Z_{b}} {\rm tr}_{b}\big[e^{i (B+H_{b})t} e^{-i H_{b}t} e^{-\beta H_{b}} \big] = \frac{1}{Z_{b}}{\rm tr}_{b}\big[ e^{-i H_{b}t}e^{i (B+H_{b})t} e^{-\beta H_{b}} \big]  \simeq  \langle \hat{O}_{3}(t) \rangle  =  e^{-\frac{g w}{8}t^{2} (t^{2} + 4)\coth \frac{\beta w}{2}}. 
\end{equation}

It follows that the expression for the rate of energy transfer in Eq.~(\ref{eq:transrate}) for the case of  Harmonic oscillator baths coupled to the chromophores can be approximated as 
\begin{eqnarray}
	\label{eq:harmonicrate}
	k(t) & \simeq & 2\:{\rm Re}\:\bigg\{p J_{1}^{2}\:\int_{0}^{t}dt' e^{i(\epsilon_{A}-\epsilon_{D_{1}})(t-t')}\langle O_{2}(t,t')\rangle\langle O_{1}(t,t')\rangle \nonumber \\
	&& + (1-p)J_{2}^{2}\:\int_{0}^{t}dt' e^{i(\epsilon_{A}-\epsilon_{D_{2}})(t-t')}\langle O_{2}(t,t')\rangle\langle O_{1}(t,t')\rangle \nonumber \\
	&& + e^{i\phi} \sqrt{p} \sqrt{1-p} J_{1}J_{2}\:\int_{0}^{t}dt' e^{i(\epsilon_{A}-\epsilon_{D_{1}})t}e^{-i(\epsilon_{A}-\epsilon_{D_{2}})t'}\langle O_{3}(t')\rangle\langle O_{3}(t)\rangle\langle O_{1}(t,t')\rangle \nonumber \\
	&& + e^{-i\phi} \sqrt{p} \sqrt{1-p} J_{2}J_{1}\:\int_{0}^{t}dt' e^{i(\epsilon_{A}-\epsilon_{D_{2}})t}e^{-i(\epsilon_{A}-\epsilon_{D_{1}})t'}\langle O_{3}(t')\rangle\langle O_{3}(t)\rangle \langle O_{1}(t,t')\rangle\bigg\}.
\end{eqnarray}

\subsubsection{Numerical Investigations}

\begin{figure}[!htb]
	\resizebox{12 cm}{7.5 cm}{\includegraphics{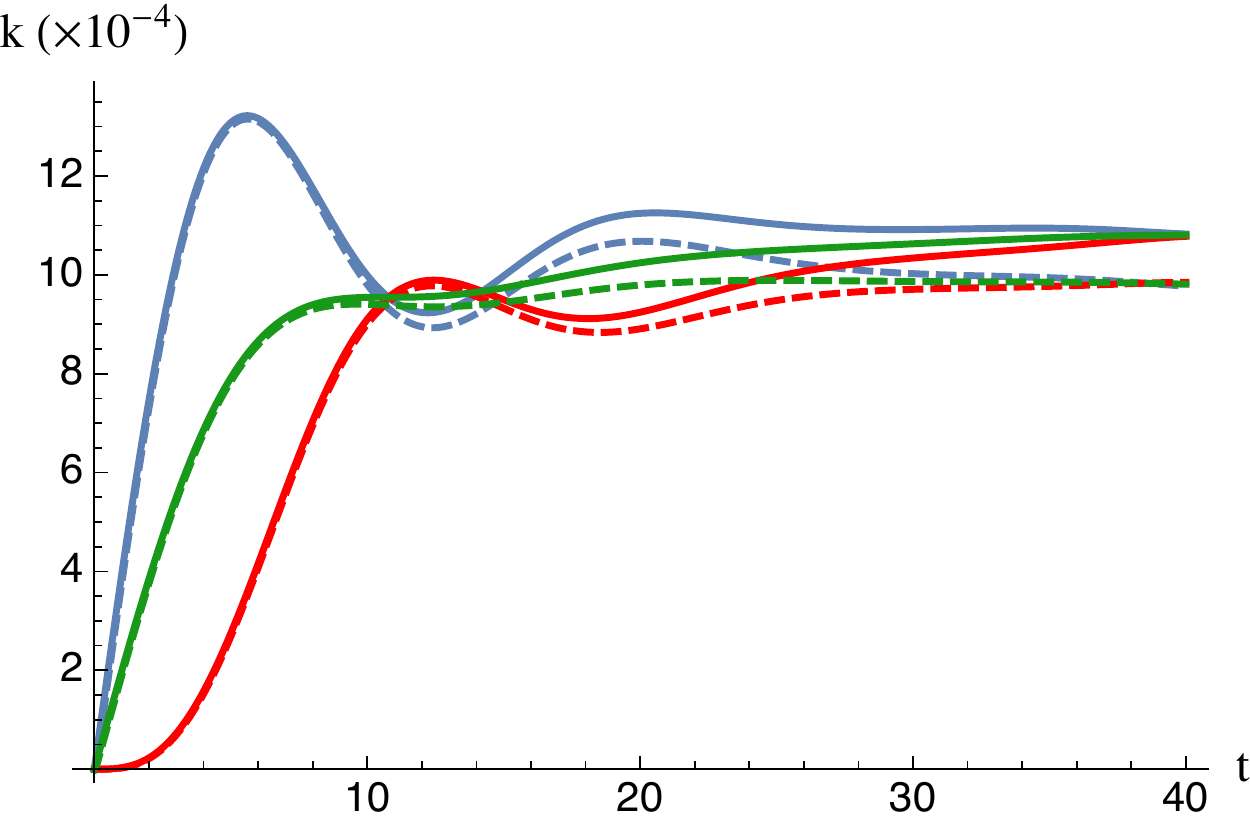}}
	\caption{Comparison between the energy transfer rates obtained through direct numerical integration of the evolution equation for the system of three chromophores - two donors and one acceptor - with each chromophore coupled to independent harmonic oscillators (with Hilbert space dimension truncated to 4 for each harmonic oscillator). The dashed lines show the exact evolution when the initial state of the donors is as given in Eq.~(\ref{eq:IniState}) with $p=1/2$ and $\phi=0$ (blue), $\phi=\pi/2$ (green) and $\phi=\pi$ (red). The solid lines of the same colors are the corresponding transfer rates in arbitrary units as given in Eq.~(\ref{eq:transrate}). We have assumed $J_{1} = J_{2} = J = 0.01$ as mentioned earlier in the text. Note that when $\phi=\pi/2$ with $J_{1}  = J_{2}$, Eq.~(\ref{eq:transrate}) is equivalent to the energy transfer rate equation for two donors independently interacting with the acceptor with no coherences between them. The comparison between the predictions of Eq.~(\ref{eq:transrate}) and the results from direct numerical integration gives serves as a test for the validity of Eq.~(\ref{eq:transrate}), which in turn can be extended to non-mesoscopic environments.\label{fig1}}. 
\end{figure}

For numerical integration of the exact evolution equations for the system along with the harmonic oscillator baths attached to each chromophore, we had to restrict the Hilbert space of the oscillators to a few dimensions. First we did a comparison of the energy transfer rate obtained through direct integration of the whole system with the rate obtained from the expression in Eq.~(\ref{eq:transrate}) for different values of $\phi$ with $p=1/2$ as shown in Fig.~(\ref{fig1}). Numerical evaluation of the integrals in Eq.~(\ref{eq:transrate}) for each value of $t$ was done using the standard trapezoidal integration routines available in {\em matlab} with the traces over the environment evaluated using the closed form expressions obtained in Eqs.~(\ref{eq:wig1}), (\ref{eq:wig2}) and (\ref{eq:wig3}).

\begin{figure}[!htb]
	\resizebox{12 cm}{7.5 cm}{\includegraphics{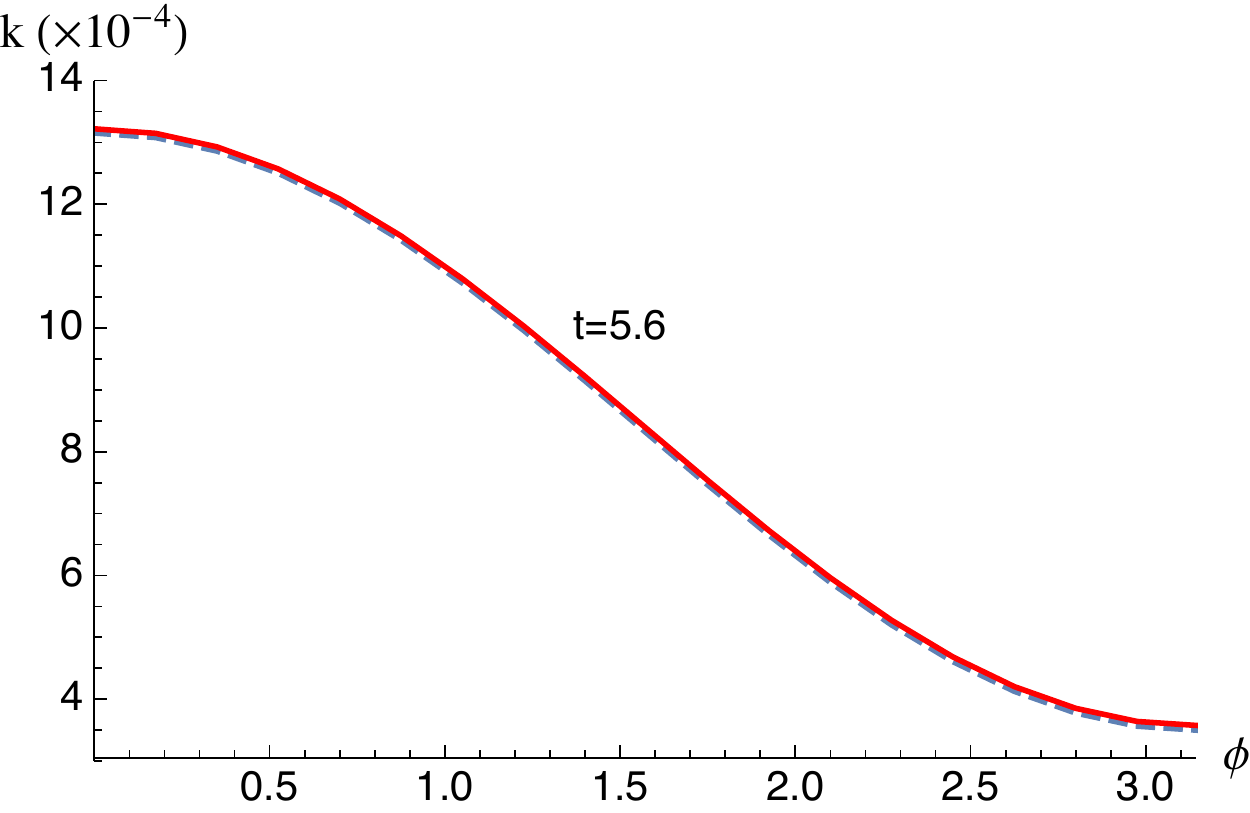}}
	\caption{Comparison between the energy transfer rates obtained through direct numerical integration of the evolution equation for the system of three chromophores - two donors and one acceptor - with each chromophore coupled to independent harmonic oscillators (with Hilbert space dimension truncated to 4 for each harmonic osscilator). The blue dashed line shows the dependence of the energy transfer rate at time $t=5.6$ (in arbitrary units) on the relative phase $\phi$ of the initial superposition state of the two donors given in Eq.~(\ref{eq:IniState}) as computed using exact numerical unitary evolution. The red line shows the same dependence as computed using Eq.~(\ref{eq:transrate}).  Note that for the system considered in~\cite{Jang:JPhysChemB:2007}, $t=5.6$ in the units used here correspond roughly to 10 femtoseconds. \label{fig2}}. 
\end{figure}

The comparison between the numerically computed rate and Eq.~(\ref{eq:transrate}) for different values of $\phi$ corresponding to a particular value of $t$ is shown in Fig.~\ref{fig2}. From Figs.~\ref{fig1} and \ref{fig2} we see that there is good agreement between the rate expression we obtained and the exact rates giving us further confidence in using these expression even in those cases where the environment is not mesoscopic and exact numerical integration is not possible. 

\begin{figure}[!htb]
	\resizebox{12 cm}{7.5 cm}{\includegraphics{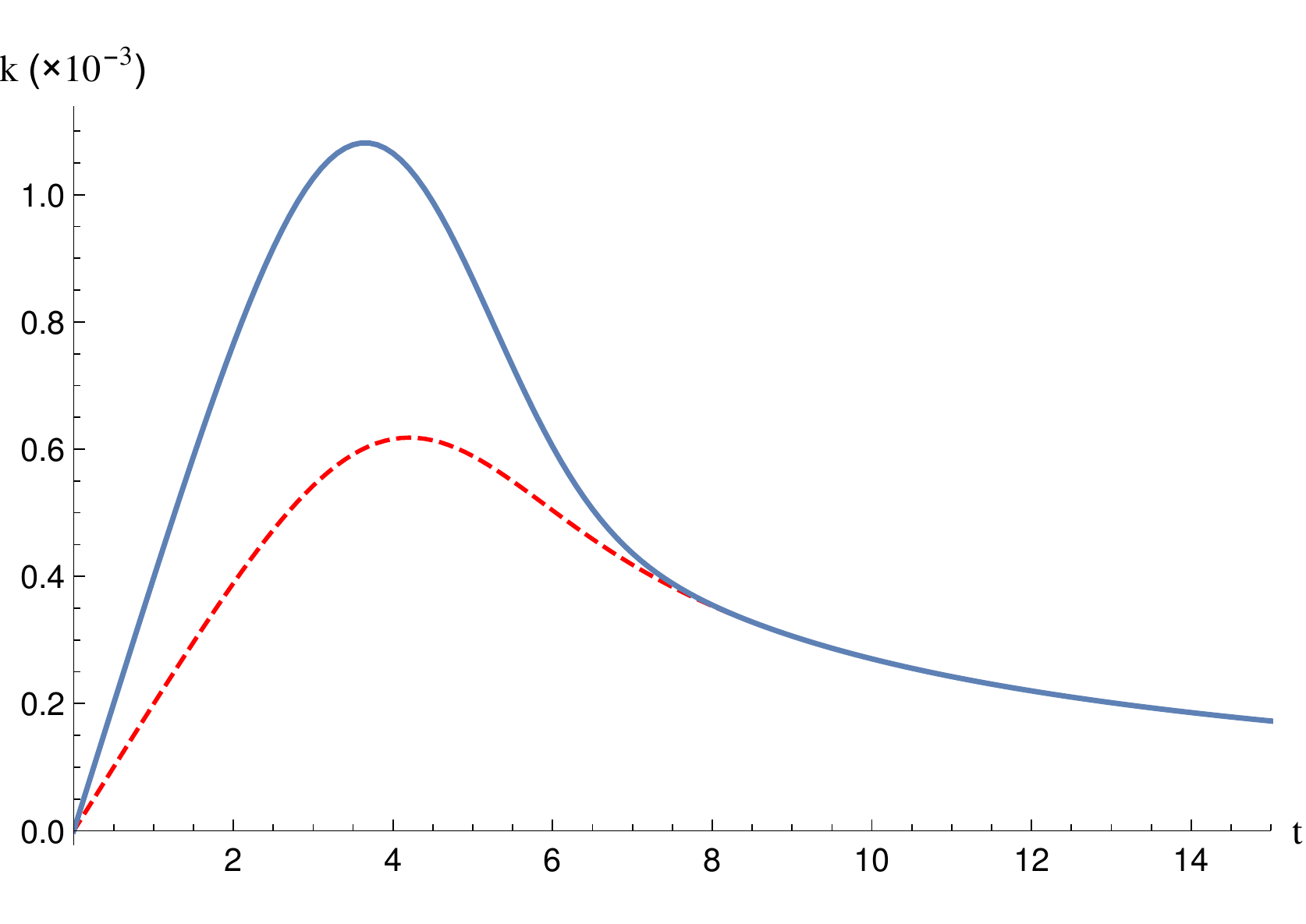}}
	\caption{Comparison of the two donor case with the single donor one: The blue (solid) line shows the energy transfer rate between two donors and an acceptor when the donors are in a coherent superposition state in the single excitation sector. The red (dashed) line shows the transfer rate between a single donor and acceptor. For both cases the donor(s) and the acceptor are coupled to Harmonic oscillator baths. The initial state of the two donors has $p=1/2$ and $\phi=0$ such that the transfer rates are given by Eq.~(\ref{eq:harmonicrate}) \label{fig3}}. 
\end{figure}

\begin{figure}[!htb]
	\resizebox{12 cm}{7.5 cm}{\includegraphics{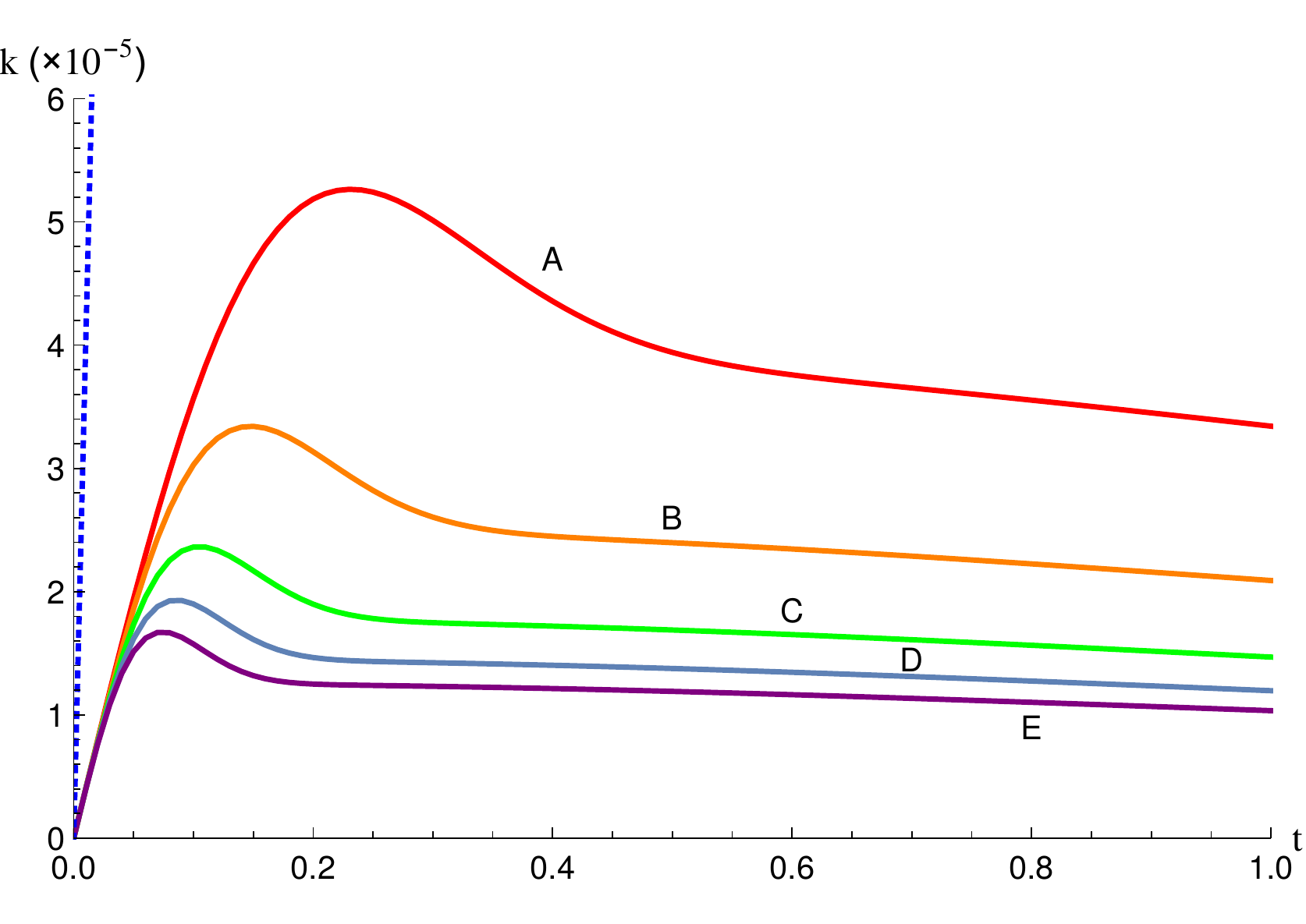}}
	\caption{The energy transfer rate between two donors and one acceptor corresponding to different temperatures of their respective single harmonic oscillator environment. The temperatures (in Kelvin) corresponding the labeled curves are are respectively A: 100, B: 250, C: 500, D: 750 and E:1000. The dotted line shows a portion of the transfer rate at $T=0$ that is plotted in Fig.~\ref{fig3} for comparison. The initial state of the two donors has $p=1/2$ and $\phi=0$ such that the transfer rates are given by Eq.~(\ref{eq:harmonicrate}) \label{fig3a}}. 
\end{figure}

In Fig.~\ref{fig3} we compare the energy transfer rate between two coherent donors in an initial state with $p=1/2$ and $\phi = 0$ in the single excitation sector with the transfer rate from a single donor also carrying a single initial excitation. We see that the coherence between the donors does indeed modulate the transfer rates at short times. For the B850-B800 system in~\cite{Jang:JPhysChemB:2007} the increased energy transfer rates is found at time scales of a few femtoseconds. This modulation can, in principle, be detected through the corresponding changes in the observed spectra at high frequencies as described earlier.  In Fig.~\ref{fig3} the energy transfer rate is computed assuming zero temperature for the harmonic oscillator bath. In Fig.~\ref{fig3a} we plot the energy transfer rate for different temperatures using the same system parameters as in Fig.~\ref{fig3}. As expected the higher temperature of the bath is seen to reduce the overall energy transfer date. More significantly we see that the initial enhancement in the rate due to the coherence between the donors vanishes more rapidly as the temperature increases and the duration for which the enhancement exists is also reduced to a fraction of a femtosecond for the example system in~\cite{Jang:JPhysChemB:2007}.

\subsection{Qubits as bath modes}

A group of $N$ qubits where $N=1,2,3,\ldots$ is taken as the environment attached to each of the three chromophores in this section. The interaction between the chromophore and its $N$ qubit environment is for the form 
\[ H_{eb} = \sum_{s,j} g_{sj}\sigma_{x}^{(j)} |s\rangle \langle s|, \qquad s=1,2,A \quad {\rm and} \quad j=1, \ldots,N,\] 
where $\sigma_{k}^{(j)}$ are Pauli matrices acting on the $j^{\rm th}$ qubit. The free evolution of the bath qubits attached to each chromophore is governed by the Hamiltonian, 
\[ H_{b} = \frac{1}{2} \sum_{j=1}^{N} \epsilon_{j} \sigma_{z}^{(j)}. \]

\begin{figure}[!htb]
	\resizebox{16.5 cm}{4.7 cm}{\includegraphics{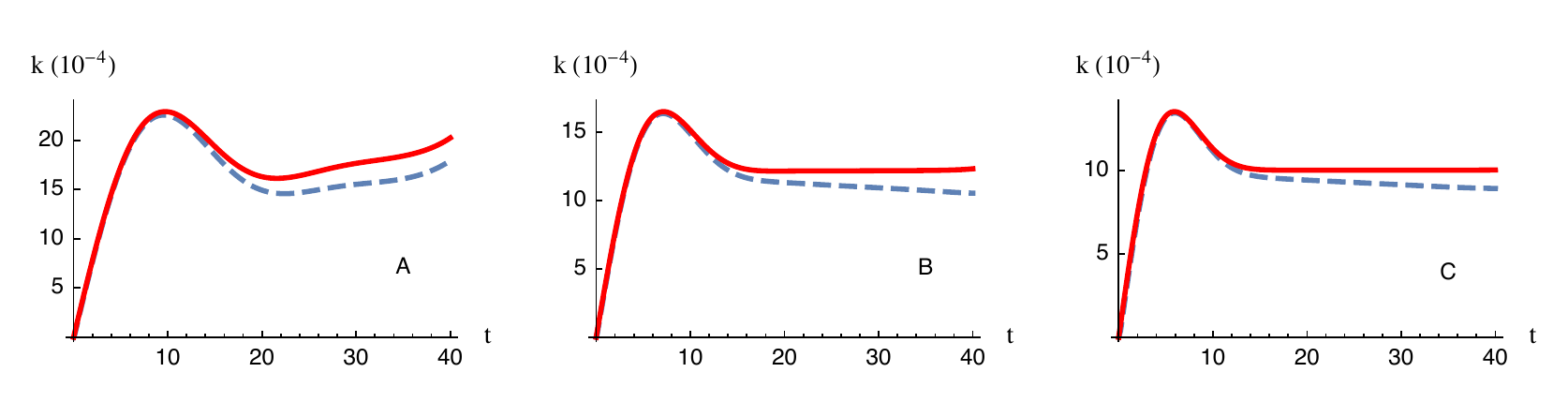}}
	\caption{ Comparison of the energy transfer rates between two donors and an acceptor with their respective environments modelled by a collection of $N$ qubits. In each of the three plots, the blue (dashed) line corresponds to the energy transfer rate computed from the exact unitary evolution of the whole system including the qubit baths while the red (solid) line corresponds to the transfer rate predicted by Eq.~(\ref{eq:transrate}). In Plot $A$, the environment of each of the three chromophores is a single qubit while in $B$ each chromophore is attached to a bath made of two qubits. Plot $C$ corresponds to baths made of three qubits each.  \label{fig4}}. 
\end{figure}

In Fig.~\ref{fig4} we plot the energy transfer rate between two donors in the initial state characterised by $p=1/2$ and $\phi=0$ and the acceptor for three cases corresponding to $N=1$, $2$ and $3$ respectively. The rate as obtained using exact unitary evolution of the entire system including the bath qubits is compared with the rate obtained using Eq.~(\ref{eq:transrate}). We again find that there is good agreement between the numerically computed transition rates and the predictions of Eq.~(\ref{eq:transrate}). As the number of qubits in the bath become larger, the agreement between the numerical and theoretical results is improved. 

\begin{figure}[!htb]
	\resizebox{16.5 cm}{13 cm}{\includegraphics{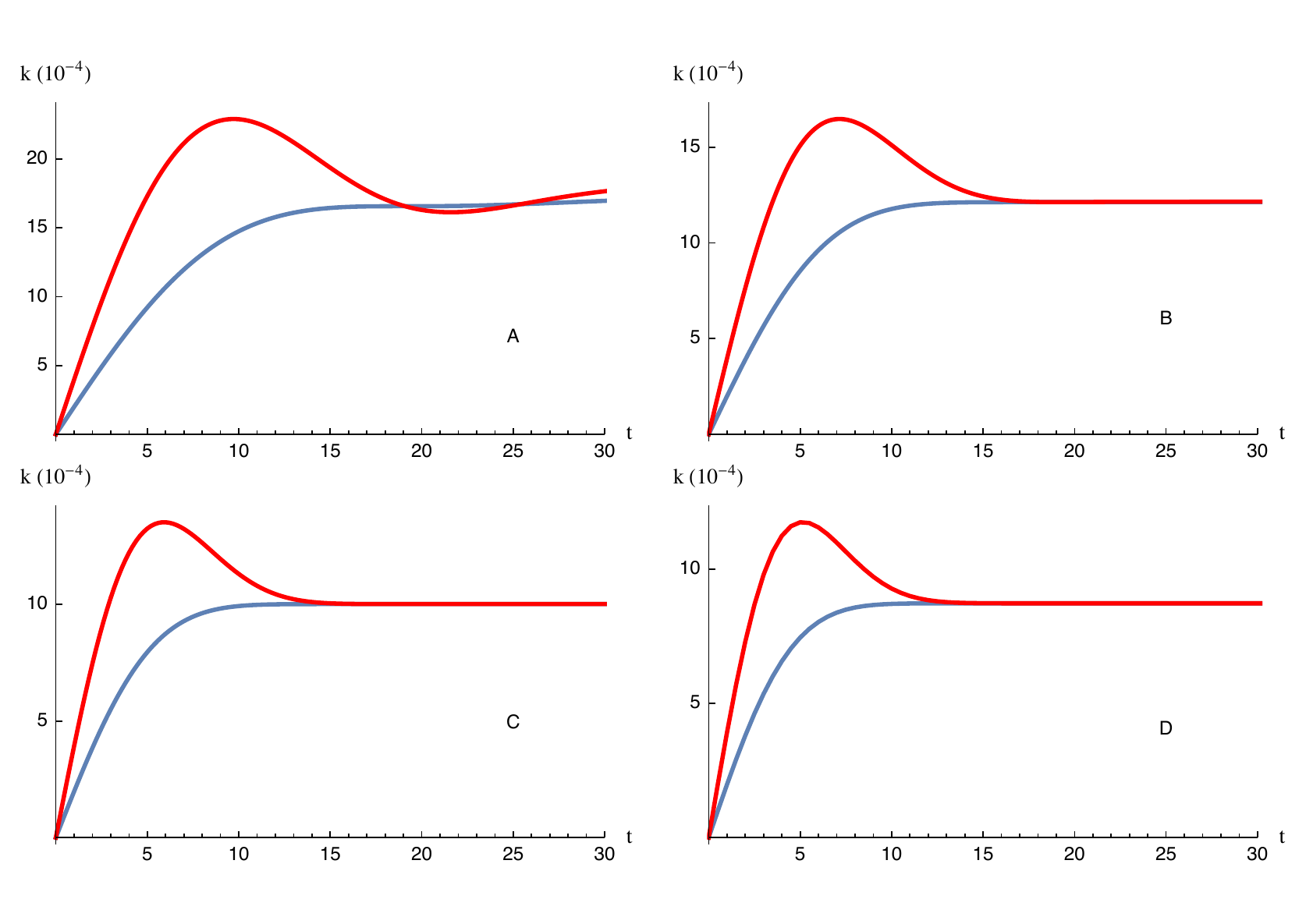}}
	\caption{ Comparison of the energy transfer rates between two donors and an acceptor in the single excitation sector with that between one donor and and an acceptor. The rates are computed using Eq.~(\ref{eq:transrate}) and the respective environments of each chromophore is modelled by a collection of $N$ qubits. In each of the three plots, the red line corresponds to the energy transfer rate for the two donor case with the donors in an initial state with $p=1/2$ and $\phi=0$. The blue line corresponds to the singe donor case. In Plot $A$, the environment of each of the three chromophores is a single qubit while in $B$ each chromophore is attached to a bath made of two qubits. Plot $C$ corresponds to baths made of three qubits each and plot $D$ is for four qubit baths.  \label{fig5}}. 
\end{figure}

In Fig.~\ref{fig5} we plot the compare the energy transfer rates as computed using Eq.~(\ref{eq:transrate}) for two donors in the single excitation sector with that for a single donor. The comparison is done for four different choices of the baths corresponding to 1, 2, 3 and 4 qubits respectively coupled to each of the chromophores. We find that as the number of qubits in the bath is increased the asymptotic transfer rates, after the initial coherence between the donors has vanished, matches better. For a very small bath, the dynamics of the bath itself has a significant effect on the transfer rate through the trace terms in Eq.~(\ref{eq:transrate}) even at long times. Again we see that there is a modification of the energy transfer rate at short times that is attributable to the coherences between the donors that persists independent of the nature and dimensionality of the environments affecting each of the chromophores. 

\section{Discussion \label{conclusion}}

In many excitonic energy transfer processes, especially biologically relevant ones, the energy donors are located in close proximity to each other so that a single exciton being delocalised across many chromophores is a very realistic possibility~\cite{lambert_quantum_2013,fassioli_photosynthetic_2014}. The donors themselves are excited typically by absorbing a photon. While the ``size'' of a photon itself may not be a well posed question, by most estimates the extent of the photon that is absorbed can span several of these donor chromophores. So it is reasonable in many scenarios to assume that more than one of the donor chromophores may be excited into a joint superposition state in the single exciton sector by the incoming radiation. Our investigations are aimed at capturing the essential features of the onward transfer of the energy absorbed by the donors to an acceptor through FRET like mechanisms. In particular we are interested in the modifications to the energy transfer rate when there is quantum coherence between the donors. We also studied the effect of mesoscopic environments on the transfer rate. 

We find that the coherence between the donors can lead to both enhanced or reduced energy transfer rates at short times relative to the rate when there is only one donor. In the arbitrary units we have used, `short times' refers to intervals that are short relative to the time scale set by the inverse of the coupling strengths $J_{i}$ between the donors and acceptors. Meanwhile for the prototype system discussed in~\cite{Jang:JPhysChemB:2007} this means a few femtoseconds. The type of modulation of the rate depends on the amplitudes and relative phase of the initial superposition state of the two donors. We are able connect the modification in the transfer rate to observable spectral features in the stimulated emission profile of the donors. It must however be noted that the stimulated emission profile also assumes an initial superposition state which means that the spectrum must be measured in the presence of a radiation field that is identical to the one that is used in the energy transfer process. 

In our investigations the environment of each of the chromophores had relatively passive roles to play. They were primarily responsible for removing the coherences between the donors and making the energy transfer rate in the two donor case identical to the one donor case at long times. Through numerical computation we find that for both harmonic oscillator and qubit baths, the behaviour of the energy transfer rate as a function of time qualitatively shows the same features. In photosynthesis and related energy transfer processes, it is suspected that the environment of the chromophores may play a more active role in both facilitating and enhancing efficient energy flow from the donors to the acceptors\cite{sarovar_quantum_2010,huelga_vibrations_2013}. Addressing this possibility in the context of our analytic and numerical results remains to be done.

\appendix
\section{Rate expression for two coherently excited donors \label{appendixA}}

In the interaction picture, the probability that and initial state of two donors, an acceptor and their respective environments of the form 
\[ \rho(0) = \frac{1}{Z_{b}} |\psi \rangle e^{-\beta H_{b}} \langle \psi |, \qquad |\psi\rangle = \sqrt{p} |D_{1} \rangle + e^{-i\phi} \sqrt{1-p} |D_{2} \rangle,\]
with the single excitation localized in the donor chromophores transitions to the state $|A\rangle$  is given by 
\begin{equation}
	\label{eq:appA1}
	p_{A}(t) = {\rm tr}_{b} \big[ \langle A_{I}\vert U_{I}(t,0)\rho_{I}(0)U_{I}^{\dagger}(t,0)\vert A_{I} \rangle \big],
\end{equation}
with $\rho_{I}(0) = \rho(0)$.  The unitary time evolution operator in the interaction picture is
\begin{equation}
	\label{eq:appA2}
	U_{I}(t,0) = e^{-i\int_{0}^{t}dt'V_{I}(t')}, \qquad V_{I}(t) = e^{iH_{0}t} V e^{-iH_{0}t}, 
\end{equation}
where $H_{0}$ and $V$ are given in Eqs.~(\ref{eq:h0}) and (\ref{eq:perturb}) respectively and
\[ \vert A_{I}(t)\rangle = e^{i H_{0}t}\vert A\rangle = e^{i(\epsilon_{A}+B_{A}+H_{b})t}\vert A\rangle. \]
Treating $V$ as a perturbation for small values of the couplings $J_{1}$ and $J_{2}$, we can do a series expansion for the unitary operator and consider the first few terms:
\begin{equation}
	\label{eq:appA3}
	U_{I}(t,0) \simeq \mathbb{I} -i\int_{0}^{t}dt'V_{I}(t') -\int_{0}^{t}dt'\int_{0}^{t}dt''V_{I}(t')V_{I}(t'') + \ldots
\end{equation}
Inserting (\ref{eq:appA3}) into (\ref{eq:appA1}, we find that the leading order non-vanishing term is of order two in the coupling constants $J_{1}$ and $J_{2}$ and is given by
\begin{equation}
	\label{eq:appA4}
	p_{A}(t)  \simeq   \frac{1}{Z_{b}}\int_{0}^{t} dt' \int_{0}^{t}dt'' \, {\rm tr}_{b} \big[ \langle A| e^{-iH_{0}t} e^{i H_{0}t'} V e^{-i H_{0}t'} |\psi \rangle e^{-\beta H_{b}} \langle \psi |  e^{i H_{0}t''} V e^{-i H_{0}t''} e^{iH_{0}t}| A\rangle \big].
\end{equation}
Using
\[ e^{iH_{0}t}|A\rangle = e^{i(\epsilon_{A}+B_{A}+H_{b})t}| A\rangle, \]
\[   e^{-iH_{0}t}| \psi \rangle = \sqrt{p} \, e^{-i(\epsilon_{1} + B_{1} + H_{b} )t} |D_{1} \rangle + \sqrt{1-p} \,e^{-i\phi} e^{-i(\epsilon_{2} + B_{2} + H_{b}  )t } |D_{2} \rangle, \]
and
\[ V|D_{1}\rangle = J_{1} |A\rangle, \qquad V|D_{2}\rangle = J_{2} |A\rangle, \]
we obtain
\begin{eqnarray}
	\label{eq:appA5}
p_{A}(t) & \simeq &  \frac{1}{ Z_{b}}\int_{0}^{t} dt' \int_{0}^{t} dt'' \, {\rm tr}_{b} \big\{  e^{-i(\epsilon_{A}+B_{A}+H_{b})t} e^{i(\epsilon_{A}+B_{A}+H_{b})t'} \nonumber \\
&& \qquad \times \; \big[J_{1} \sqrt{p} \, e^{-i(\epsilon_{1} + B_{1} + H_{b} )t'} + J_{2} \sqrt{1- p} \, e^{-i\phi} e^{-i(\epsilon_{2} + B_{2} + H_{b} )t'}  \big] \nonumber  \\
	&& \qquad \times \, e^{-\beta H_{b}} \big[J_{1 }\sqrt{p} \, e^{i(\epsilon_{1} + B_{1} + H_{b} )t''} + J_{2} \sqrt{1-p} \, e^{i\phi}e^{i(\epsilon_{2} + B_{2} + H_{b} )t''}  \big] \nonumber \\
	&& \qquad \times \, e^{-i(\epsilon_{A}+B_{A}+H_{b})t''}  e^{i(\epsilon_{A}+B_{A}+H_{b})t} \big\}
\end{eqnarray}
The cyclic property of the trace lets us cancel the first and last terms inside curly braces in the above expression and expanding out the remaining terms  gives us Eq.~(\ref{eq:transitionprob}). The energy transfer rate is obtained by differentiating the expression for $p_{A}(t)$ under the integral sign. For taking the time derivative of $p_{A}(t)$ we have to evaluate expressions of the form
\[ G(t) = \dfrac{d}{dt}\int_{0}^{t}dt'\int_{0}^{t}dt''f(t',t'') \]
Now, this integral can be re-written as
\begin{equation}
	\label{eq:appA5a}
	G(t) = \dfrac{d}{dt}\int_{0}^{t}dt'F(t',t),
\end{equation} 
where
\[	F(t',t) = \int_{0}^{t}dt''f(t',t''), \]
The Leibniz formula for differentiating under the integral sign is, 
\[ \frac{d}{dx}\int_{y_{1}(x)}^{y_{2}(x)}dx'f(x,x') = f(x,y_{2})\frac{dy_{2}(x)}{dx} - f(x,y1)\frac{dy_{1}(x)}{dx} + \int_{y_{1}(x)}^{y_{2}(x)}dx'\dfrac{\delta}{\delta x} f(x,x'). \]
Applying the Leibniz formula to Eq.~(\ref{eq:appA5a}) we get 
\[ G(t) =F(t,t)+ \int_{0}^{t}dt'\dfrac{d}{dt}F(t',t) =  \int_{0}^{t}dt''f(t,t'') + \int_{0}^{t}dt'\dfrac{d}{dt}\int_{0}^{t}dt''f(t',t'') \] 
Using the Leibniz formula once again to evaluate the derivative in the second term  we get,
\begin{equation}
	\label{eq:appA5b}
	G(t) =\int_{0}^{t}dt''f(t,t'') + \int_{0}^{t}dt'f(t',t) = \int_{0}^{t}dt'f(t,t') + \int_{0}^{t}dt'f(t',t)
\end{equation}
where we have relabelled the dummy variable $t''$ to $t'$ in one of the terms. Using Eq.~(\ref{eq:appA5b}) to take the time derivative of the transition probability (\ref{eq:transitionprob}), we obtain,
\begin{eqnarray}
\label{eq:appA6}
k(t) & = &  \dfrac{d}{dt}p_{A}(t) \nonumber \\
& = & p \frac{ J_{1}^{2}}{ Z_{b}}  \int_{0}^{t}  dt' \, e^{i\epsilon_{A}(t-t')} e^{-i\epsilon_{1}(t-t')}\:{\rm tr}_{b} \big[ e^{i(B_{A}+H_{b})(t-t')}e^{-i(B_{1}+H_{b})t}e^{-\beta H_{b}}e^{i(B_{1}+H_{b})t'}\big]\nonumber \\
&& + p \frac{ J_{1}^{2}}{ Z_{b}}  \int_{0}^{t}  dt' \, e^{i\epsilon_{A}(t'-t)} e^{-i\epsilon_{1}(t'-t)}\:{\rm tr}_{b} \big[ e^{i(B_{A}+H_{b})(t'-t)}e^{-i(B_{1}+H_{b})t'}e^{-\beta H_{b}}e^{i(B_{1}+H_{b})t}\big]\nonumber \\
&&  +  (1-p) \frac{J_{2}^{2}}{ Z_{b}} \int_{0}^{t}  dt'  \, e^{i\epsilon_{A}(t-t')}  e^{-i\epsilon_{2}(t-t')} \: {\rm tr}_{b} \big[ e^{i(B_{A}+H_{b})(t-t')} e^{-i(B_{2}+H_{b})t}e^{-\beta H_{b}}e^{i(B_{2}+H_{b})t'}\big] \nonumber \\
&&  +  (1-p) \frac{J_{2}^{2}}{ Z_{b}} \int_{0}^{t}  dt'  \, e^{i\epsilon_{A}(t'-t)}  e^{-i\epsilon_{2}(t'-t)} \: {\rm tr}_{b} \big[ e^{i(B_{A}+H_{b})(t'-t)} e^{-i(B_{2}+H_{b})t'}e^{-\beta H_{b}}e^{i(B_{2}+H_{b})t}\big] \nonumber \\
&& +  e^{i\phi} \sqrt{p} \sqrt{1-p} \frac{J_{1}J_{2}}{ Z_{b}}  \int_{0}^{t} dt'  \,  e^{i\epsilon_{A}(t-t')} e^{-i\epsilon_{1}t}e^{i\epsilon_{2}t'} \: {\rm tr}_{b} \big[ e^{i(B_{A}+H_{b})(t-t')} e^{-i(B_{1}+H_{b})t} e^{-\beta H_{b}} e^{i(B_{2}+H_{b})t'} \big] \nonumber \\
&& +  e^{i\phi} \sqrt{p} \sqrt{1-p} \frac{J_{1}J_{2}}{ Z_{b}}  \int_{0}^{t} dt'  \,  e^{i\epsilon_{A}(t'-t)} e^{-i\epsilon_{1}t'}e^{i\epsilon_{2}t} \: {\rm tr}_{b} \big[ e^{i(B_{A}+H_{b})(t'-t)} e^{-i(B_{1}+H_{b})t'} e^{-\beta H_{b}} e^{i(B_{2}+H_{b})t} \big] \nonumber \\
&& + e^{-i\phi} \sqrt{p} \sqrt{1-p}  \frac{J_{2}J_{1}}{ Z_{b}}  \int_{0}^{t}  dt'  \, e^{i\epsilon_{A}(t-t')} e^{-i\epsilon_{2}t}e^{i\epsilon_{1}t'} \:  {\rm tr}_{b} \big[ e^{i(B_{A}+H_{b})(t-t')} e^{-i(B_{2}+H_{b})t}e^{-\beta H_{b}}e^{i(B_{1}+H_{b})t'} \big]\nonumber \\
&& + e^{-i\phi} \sqrt{p} \sqrt{1-p}  \frac{J_{2}J_{1}}{ Z_{b}}  \int_{0}^{t}  dt'  \, e^{i\epsilon_{A}(t'-t)} e^{-i\epsilon_{2}t'}e^{i\epsilon_{1}t} \:  {\rm tr}_{b} \big[ e^{i(B_{A}+H_{b})(t'-t)} e^{-i(B_{2}+H_{b})t'}e^{-\beta H_{b}}e^{i(B_{1}+H_{b})t} \big].\nonumber\\ 
\end{eqnarray}
We notice that the first two terms in the above expression are complex conjugates of each other. Similarly, third and fourth, fifth and eighth and sixth and seventh terms are also complex conjugate pairs, leading to Eq.~(\ref{eq:transrate}). Note that in Eq.~(\ref{eq:transrate}) the traces over the three sets of mutually decoupled bath modes associated with the two donors and the acceptor respectively have been further separated out. 

\acknowledgements

Sreenath K M acknowledges the support of the Department of Science and Technology, Government of India, through the INSPIRE fellowship scheme (No.~DST/INSPIRE-SHE/IISER-T/2008). Anil Shaji acknowledges the support of the Department of Science and Technology, Government of India, through the Ramanujan Fellowship program (No. SR/S2/RJN- 01/2009).

\bibliography{TwoDonorOneAcceptor}

\end{document}